\pdfoutput=1 
\documentclass{JINST}

\title{A Study of Muon Collider Background Rejection Criteria in Silicon Vertex and Tracker Detectors}

\author{V.~Di Benedetto$^a$, C.~Gatto$^b$, A.~Mazzacane$^c$, N.V.~Mokhov$^c$,
S.I.~Striganov$^c$, N.K.~Terentiev$^d$\thanks{Corresponding author.}\\
\llap{$^a$}Instituto Nazionale di Fisica Nucleare(INFN), Universita del Salento, Lecce, Italy\\
\llap{$^b$}Instituto Nazionale di Fisica Nucleare (INFN), Sezione di Napoli, 
Complesso Universita di Monte, SantAngelo, via, I-80126 Naples, Italy\\
\llap{$^c$}Fermi National Accelerator Laboratory, P.O.Box 500, Batavia, Illinois 60510, USA\\
\llap{$^d$}Carnegie Mellon University, 5000 Forbes Avenue, Pittsburgh, Pennsylvania 15213, USA\\
E-mail: \email{teren@fnal.gov}}

\abstract{The hit response of  silicon  vertex and tracking  detectors  to
muon collider beam background and  results of a study of  hit   
reducing techniques are  presented. The background caused by decays of 
the 750 GeV$/c$ $\mu^{+}$ and $\mu^{-}$ beams was simulated using the  MARS15  
program,  which  included  the  infrastructure  of  the beam  line  elements  
near  the  detector  and  the $10^{\circ}$ nozzles that  shield  the  detector  
from  this background. The ILCRoot framework, along with the Geant4 program, 
was used to simulate the hit  response  of  the  silicon vertex  and  tracker  
detectors to  the  muon decay  background remaining  after the  shielding  
nozzles. The  background  hit  reducing  techniques include timing,  
energy  deposition,  and  hit  location correlation  in  the  double  layer  
geometry.}


\begin{document}

\section{Introduction}\label{sec:sec1}

\paragraph{}The latest results \cite{M1} of a comprehensive  study of interaction 
region (IR) and  machine-detector interface (MDI)   designs   for   1.5  TeV   
muon   collider\,\cite{Gallardo,Summers,Stratakis}   demonstrate  that   the   muon   beams background  can  be  
suppressed more  than  three  orders  of  magnitude  by using properly designed
shielding cones (see  details  in \cite{S1}).  Data  was  
obtained  with the MARS15  simulation  program   \cite{MARS15}. They represent the list of background particles with their 
characteristics given at the detector surface of the MDI (two 10$^\circ$  shielding
nozzles near the interaction point (IP)).
 
The MARS15 output data were used as input for simulation of the detector hit
response in the ILCRoot framework \cite{ILCRoot} . In this paper we present 
results of ILCRoot simulation of silicon vertex and  tracking  detectors hit
response  to  the  muon  beam  background.    The  background    
reducing techniques were  studied on the  hit level. They include use of 
timing, energy deposition and hit spatial correlation in the double layer 
geometry of silicon vertex and tracking detectors.

Event tracks come from the IP.  Background comes from the beamline, most of which is not at the IP.

\section{The MARS15 Modeling Results}\label{sec:sec2}

\begin{figure}[ht]
\centering
\includegraphics[height=60mm]{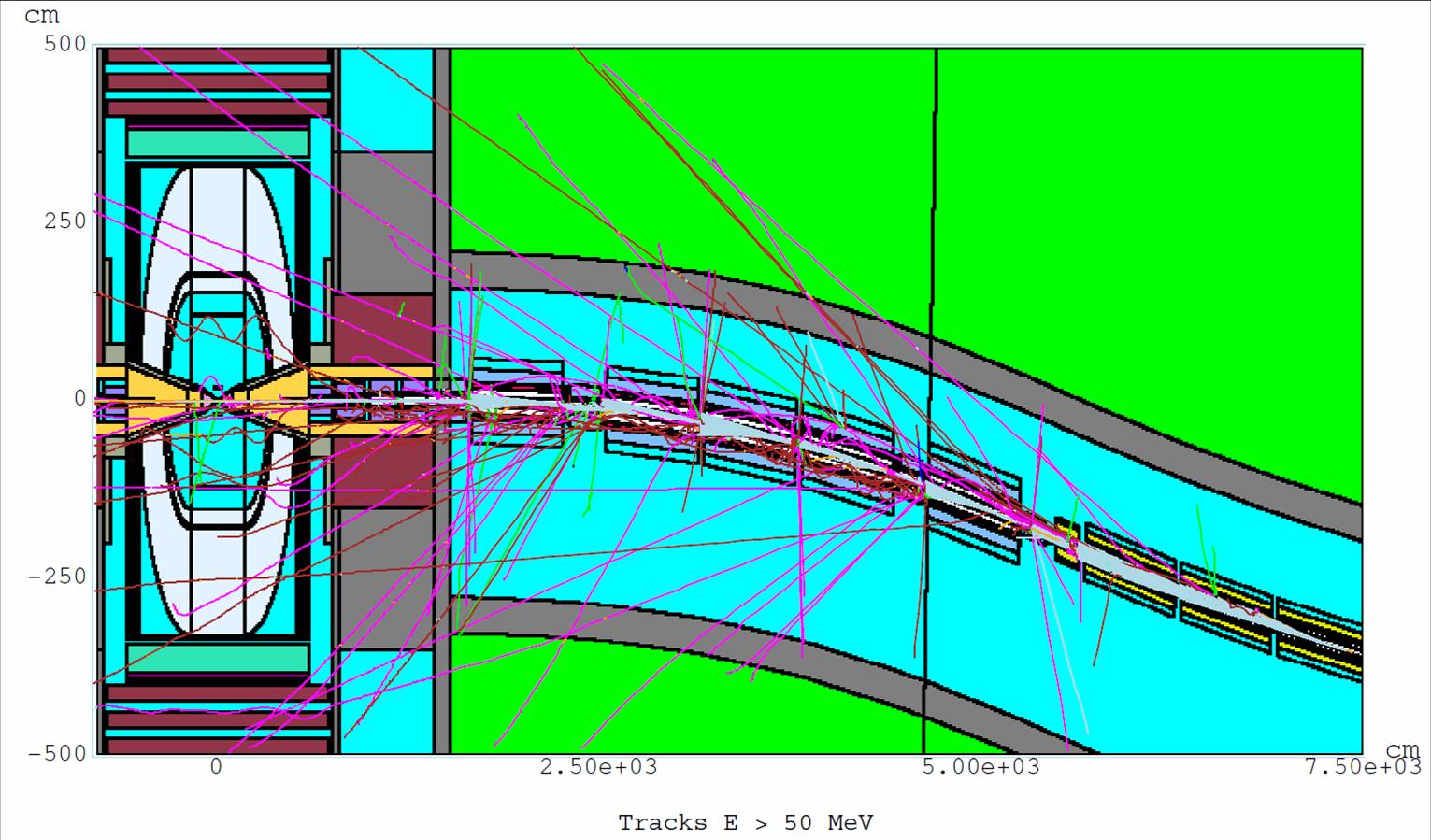}
\caption{A MARS15 model of the IR and detector with particle tracks > 1 GeV 
(mainly muons) for several forced decays of both beams.}
\label{fig:ir}
\end{figure}

\begin{figure}[ht]
\centering
\begin{minipage}{.45\textwidth}
  \centering
  \includegraphics[height=50mm]{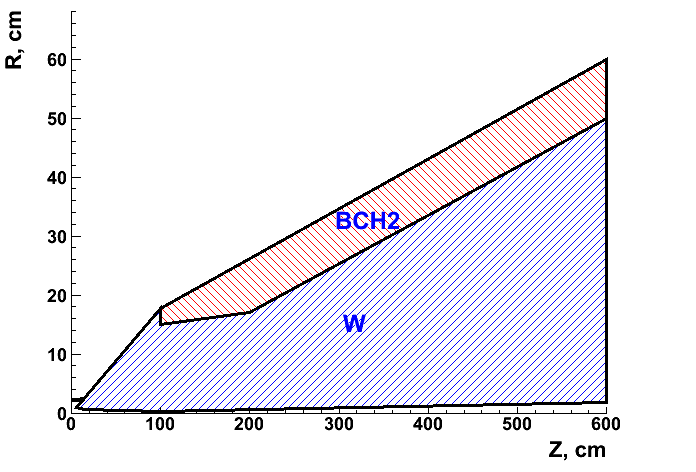}
  \caption{The shielding nozzle, 
           general RZ view (W - tungsten,  
           BCH2 - borated  polyethylene)}
  \label{fig:nzl_gen}
\end{minipage} \hfill
\begin{minipage}{.45\textwidth}
  \centering
  \includegraphics[height=45mm]{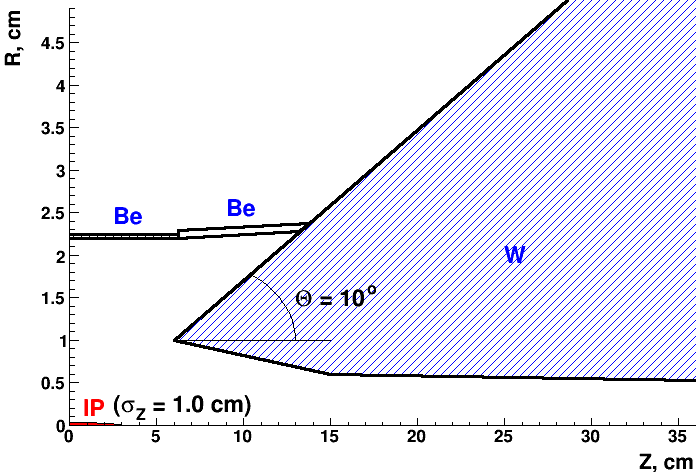}
  \caption{The shielding nozzle,
           zoom in near IP (Be - beryllium)}
  \label{fig:nzl_zoom}
\end{minipage}
\end{figure}

\paragraph{}
The major source of the detector background in $\mu^{+}$ $\mu^{-}$  
collider is the electrons 
and positrons from beam muon decays.   For 750 GeV muon beam with intensity of 
$2 \times 10^{\,12}$ per bunch there are about $4 \times 10^{\,5}$ decays  per meter per bunch 
crossing.
The decay $e^{+}$ and $e^{-}$   produce high intensity secondary particle 
fluxes in the 
beam line components and accelerator tunnel in the vicinity of the detector 
(interaction region IR, Figure~\ref{fig:ir}). 
As it was  shown in the  recent study \cite{M1},  the appropriately  designed  
interaction  region  and   machine  detector  interface  (including  shielding 
nozzles,  Figure~\ref{fig:nzl_gen} and Figure~\ref{fig:nzl_zoom} )  can 
 provide  the  reduction  of  muon  beam
  background
 by  more  than  three orders  of  magnitude  for a  muon  collider  with  a
collision  energy  of  1.5 TeV. These  results  were obtained with the MARS15 
simulation code, the framework for simulation of particle transport
and  interactions  in accelerator, detector and  shielding components.  
The  MARS15 model takes into  account  all  the  related  details  of  
geometry,  material  distributions  and  magnetic  fields  for 
collider  lattice  elements  in  the  vicinity of the detector including 
shielding nozzles.

The amount of MARS15 simulated data was limited to 4.6\% of the 
$\mu^{+}$ $\mu^{-}$ decays on the 26~m beam length yielding total of $14.6 \times10^{\,6}$  background particles 
per bunch crossing (BX). The corresponding statistical weight ($\sim$22.3) was  
taken  into  account  in  the following  ILCRoot  simulation.  For each given  
MARS15 particle its momentum was smeared azimuthally 22-23 times to get 100\% 
statistics and provide  total  yield  of $3.24 \times10^{\,8}$ particles  entering  
the detector  in  the ILCroot  simulation. The  most abundant background  consists  of  photons and  neutrons.  Table~\ref{tab:yld}  lists  these  background  
yields together  with  kinetic  energy thresholds  used  in  the MARS15 simulation
for  different  types  of particles.
\begin{table}
    \caption{The MARS15 background yields/bunch on $10^{\circ}$ nozzles surface
             and thresholds}
    \label{tab:yld}
    \begin{center}
      \begin{tabular}{|l|c|c|c|c|c|} \hline 
{\bf Particles} & { $\gamma$}  & {n}  & {e}  & {p, $\pi$} & {$\mu$}\\
                                      \hline
{\bf Yield/BX }  &  1.72x$10^{8}$  & 1.51x$10^{8}$  &  1.5x$10^{6}$  &  6.04x$10^{4}$  &  0.28x$10^{4}$\\
\hline
{\bf Threshold} &  100 keV  &  0.001 eV  & 100 keV  & 100 keV & 100 keV \\
                                 \hline
      \end{tabular}
    \end{center}
\end{table}

\section{The ILCRoot Simulation of the Hits in Vertex and Tracker 
Silicon Detectors}\label{sec:sec3}
ILCRoot \cite{ILCRoot} is the software Infrastructure for 
Large Colliders based on ROOT \cite{ROOT} and add-ons made for muon collider 
detectors studies. It makes use of the virtual Monte Carlo concept allowing one to 
select and load at run time different Monte Carlo models (Geant3, Geant4, 
Fluka). We used Geant4 \cite{GEANT4a} with the QGSP-BERT-HP-LIV
physics list for a better description of the neutron transport and low energy EM 
processes. The ILCRoot simulation presented here  was limited to the hit level 
only and did not include the front-end electronics response.

\subsection{Detector Layout}
In this work the geometry of the ILCRoot detector included the vertex (VXD) and 
tracker (Tracker) silicon subsystems as the only sensitive detectors. The other detectors
such as muon spectrometer, electromagnetic and hadron calorimeters were used
as material in interactions with background particles without hit simulation. 
Other non active components were the shielding nozzles, detector magnet coil 
and walls. 

The vertex and tracker silicon detector layouts are based on the SiD ILC 
concept \cite{SID}. The vertex subsystem comprises of five 
barrel layers with radii 3-14 cm and  length of 12 cm in 
Z-direction and eight end-cap disks. 
The tracker has five barrel layers with radii 20-120 cm and 
lengths of 50-320 cm in z-direction and fourteen end-cap disks.
In addition to the SiD ILC concept,  the silicon forward tracker 
detector with six end-cap disks was used to cover and improve 
tracking in the forward $\theta$ region with high hit occupancy. 

The barrel layer of vertex and tracker detectors has two sublayers, each
75 microns and 200 microns thick, respectively. To study an effect of double 
layer rejection criteria four sets of layout geometry were simulated in 
ILCRoot, with space between sublayers of 1 mm and 2 mm and detector magnetic 
fields of 3.5 T and 7 T. At each geometry two sets of data were simulated 
separately, MARS15 background and IP $\mu^{+}$ $\mu^{-}$ with momentum 
P = 0.2 - 10 GeV/c. The interaction point (IP) was smeared in Z with $\sigma$=1cm.  
The samples of IP muons served for estimate of efficiency in selection 
criteria.

\subsection{ILCRoot Simulation Results}
The ILCRoot simulation output data present ROOT files with records containing 
information about GEANT4 hits and tracks producing them. The hit 
is defined for each step of the particle tracking in the sensitive volume of 
detector. It has X, Y, Z coordinates, time and momentum P components of the 
track at the begining and at the end of the step, energy deposition in the step, 
particle ID etc. ILCRoot keeps detailed information about hits provided by 
GEANT4 including status of the track (if the track entered or left a sensitive volume or 
stopped in it). Table~\ref{tab:yldfr}  shows fractions of background particles 
making hits (directly or through secondary interactions) in all layers of 
vertex and tracker silicon detectors at magnetic field of 3.5 T and 1 mm space 
between sublayers. The data in Table~\ref{tab:yldfr} are not corrected for 
geometry acceptance of the silicon detectors. The overall fraction of MARS15 background particles making hits in VXD and Tracker is $\sim$3\%.
\begin{table}
    \caption{Fractions of MARS15 background particles making hits in silicon vertex 
             and tracker detectors}
    \label{tab:yldfr}
    \begin{center}
      \begin{tabular}{|l|c|c|c|c|c|} \hline 
{\bf Particles} & { $\gamma$}  & {n}  & {e}  & {p, $\pi$} & {$\mu$}\\
                                      \hline
{\bf Fraction, \%}  &  3.8  & 1.7  &  19.3  & 64.4  &  84.9\\
                                 \hline
      \end{tabular}
    \end{center}
\end{table}

\section{Data Analysis and Background Rejection Criteria}\label{sec:sec4}
The ILCRoot output data with hits were analyzed in stand-alone code to 
define and apply timing, energy deposition and double layer criteria for hits 
in the barrel layers of the VXD and Tracker only. In 
this analysis the group of hits for a given track in the given sensitive volume 
(silicon sublayer), which ends by a final hit when a track was leaving the volume or 
stopped in it, was handled as a hit cluster. It was used to sum the energy 
deposition per cluster and estimate the  number of pixels crossed by the track 
in the hit cluster. For cluster timing and position coordinates the average 
over hits was used. The final results were a hit cluster efficiency for IP muons
and a hit cluster surviving fraction for MARS15 background particles.

\subsection{Timing}
The MARS15 framework gives the time of flight of background particles 
calculated on the detector side surface of the shielding cone with respect 
to bunch crossing, BX. GEANT4 in ILCroot is tracking these particles through 
the detector, and takes into account the MARS15 time of flight and provides the time
of flight (TOF) for each hit in sensitive volume with reference  to BX.

In analysis we used instead TOF-T0 where T0 - time of flight of IP photon from 
interaction point IP with coordinates X=0, Y=0 and Z=0 to the point with IP muon
or MARS15 background particle hit coordinates. This compensates the difference 
in TOF for IP particles making hits in different layers of VXD and Tracker at 
different R and Z coordinates of the hit. The TOF-T0 time of the hit cluster 
was smeared with Gaussian time resolution of 0.2 ns. There is an additional 
smearing with $\sigma$$\sim$0.033 ns in ILCroot due to the Z distribution of 
the muon IP. 

\begin{figure}[h]
\centering
\begin{minipage}{.45\textwidth}
  \centering
  \includegraphics[height=50mm]{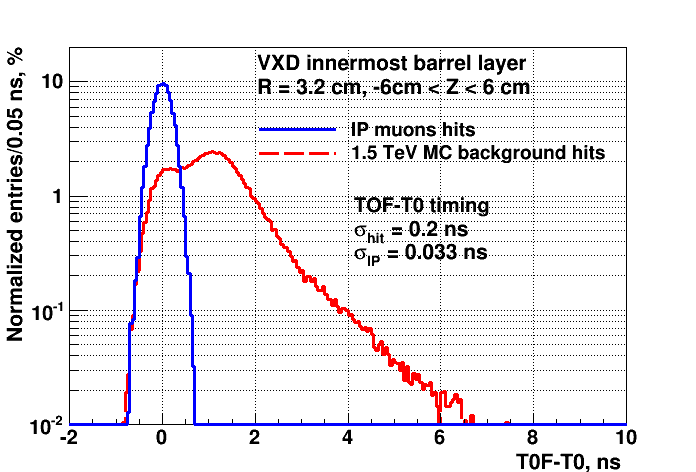}
  \caption{VXD TOF-T0 hit distribution for 
           IP muons and MARS15 background particles}
  \label{fig:vxd_ip_mars_tof-t0}
\end{minipage} \hfill
\begin{minipage}{.45\textwidth}
  \centering
  \includegraphics[height=50mm]{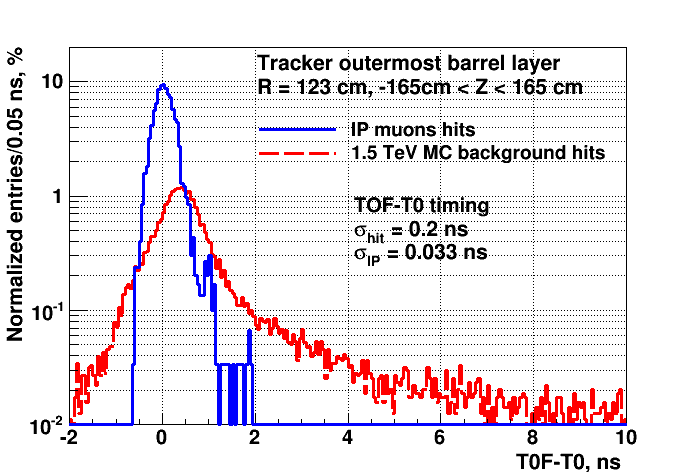}
  \caption{Tracker TOF-T0 hit distribution for 
           IP muons and MARS15 background particles}
  \label{fig:tracker_ip_mars_tof-t0}
\end{minipage}
\end{figure}

The start and width of the timing gate for TOF-T0 were defined from Gaussian fit
of the TOF-T0 distribution with conditions to have  the IP efficiency 
of $\sim$97\% in each sublayer of VXD and Tracker. Figures 
~\ref{fig:vxd_ip_mars_tof-t0}-\ref{fig:tracker_ip_mars_tof-t0} present 
TOF-T0 time distributions for hit clusters produced by IP muons and MARS15 
background particles. The backgroud distribution depends on the VXD and Tracker
layer as one can see from 
Figures~\ref{fig:vxd_ip_mars_tof-t0}-\ref{fig:tracker_ip_mars_tof-t0}, 
therefore, the conditions to keep the same IP muon efficiency in all 
layers results in different rejections of the MARS background in different 
layers.

\subsection{Energy Deposition Cut}
\begin{figure}[h]
\begin{minipage}{.45\textwidth}
  \centering
  \includegraphics[height=45mm]{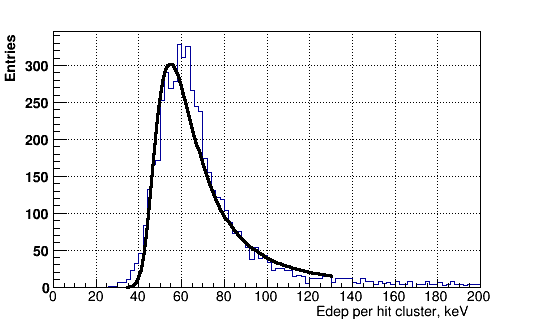}
  \caption{Energy deposition for IP muons}
  \label{fig:IP_edep_hit_cluster_Layer_18_Z_0}
\end{minipage} \hfill
\begin{minipage}{.45\textwidth}
  \centering
  \includegraphics[height=45mm]{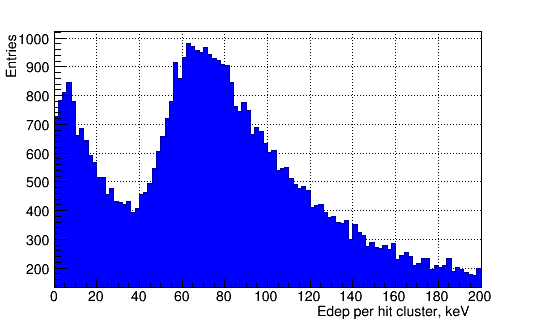}
  \caption{Energy deposition for MARS particles}
  \label{fig:MARS_edep_hit_cluster_Layer_18_Z_All}
\end{minipage}
\end{figure}

The energy deposition $E_{dep}$ of the hit cluster was defined as a sum of 
energy depositions in all hits of this hit cluster for given track in given 
sublayer of VXD and Tracker. The energy deposition resolution  $\sigma_{res}$ 
was introduced as 1/10 of Landau peak position in $E_{dep}$ distribution for IP muons. The 
cut on energy deposition (threshold) was calculated using
(Landau peak position - 2.2*$\sigma$) where $\sigma$ is Landau fit parameter 
for $E_{dep}$ distribution for IP muons. The corresponding IP muon track hit cluster efficiency per sublayer with $E_{dep}$ higher than the threshold was $\sim$96-98$\%$.
Figures~\ref{fig:IP_edep_hit_cluster_Layer_18_Z_0}-\ref{fig:MARS_edep_hit_cluster_Layer_18_Z_All} present hit cluster energy deposition for IP muons (at hit cluster Z=0) and MARS background particles (at all Z) in the Tracker outermost barrel sublayer. For fitted Landau peak position in IP muon distribution at 
$\sim$56 keV, the threshold was $\sim$42 keV with corresponding IP muon efficiency $\sim$ 98\% per sublayer. The first peak in the MARS background distribution 
corresponds to mostly $e^{-}$    resulting from background n and $\gamma$ 
interactions with silicon in any point of sensitive volume while the second peak is 
for particles crossing the sublayer.

\begin{figure}[h]
\begin{minipage}{.45\textwidth}
  \centering
  \includegraphics[height=45mm]{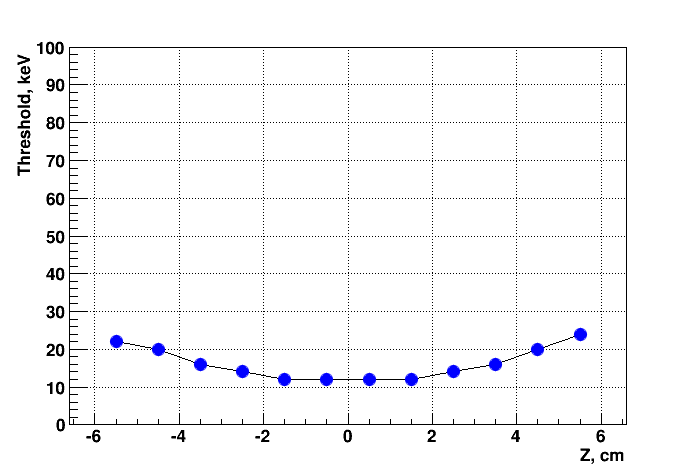}
  \caption{Energy deposition threshold in the innermost VXD barrel layer}
  \label{fig:IP_edep_hit_cluster_threshold_vs_Z_Layer_1_Z_0}
\end{minipage}\hfill
\begin{minipage}{.45\textwidth}
  \centering
  \includegraphics[height=45mm]{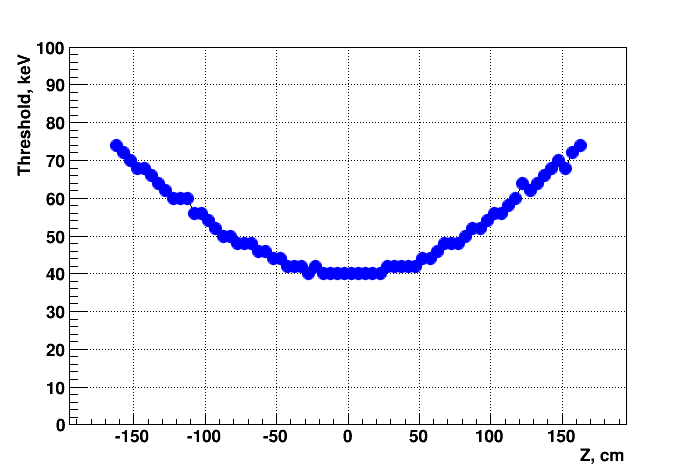}
  \caption{Energy deposition threshold in the outermost Tracker barrel layer}
  \label{fig:IP_edep_hit_cluster_threshold_vs_Z_Layer_18_Z_0}
\end{minipage}
\end{figure}

The energy deposition threshold for IP muons depends on sensitive volume 
thickness ($75\,\mu$m for VXD barrel and $200\,\mu$m for Tracker barrel sublayers)
and track polar angle ($\sim$Z position of the hit cluster in the VXD or Tracker
 barrel sublayers), see Figures~\ref{fig:IP_edep_hit_cluster_threshold_vs_Z_Layer_1_Z_0}-\ref{fig:IP_edep_hit_cluster_threshold_vs_Z_Layer_18_Z_0}.

\begin{figure}[h]
\begin{minipage}{.45\textwidth}
  \centering
  \includegraphics[height=42mm]{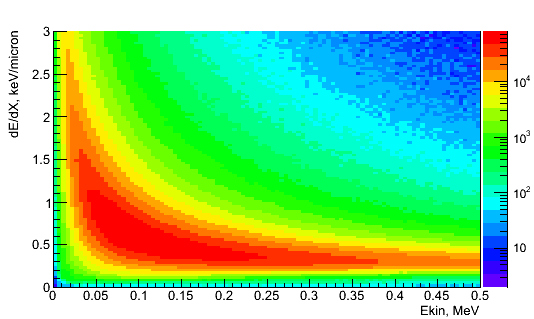}
  \caption{dE/dX vs. kinetic energy for background $e^{-}$}
  \label{fig:dedx_vs_ekin_hits_e_from_mars_g_n}
\end{minipage}\hfill
\begin{minipage}{.45\textwidth}
  \centering
  \includegraphics[height=42mm]{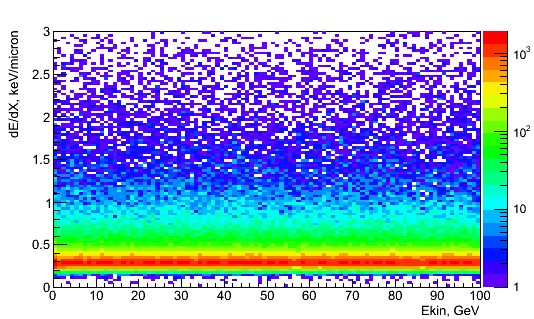}
  \caption{dE/dX vs. kinetic energy for IP muons}
  \label{fig:dedx_vs_ekin_hits_ip_mu_primary}
\end{minipage}
\end{figure}

The energy deposition selection does not provide high rejection of the muon collider background due to large dE/dX at the end of range for low energy $e^{-}$ coming from background n and $\gamma$ interactions. To estimate dE/dX we use GEANT4 energy deposition per step divided by the length of the step in the hit. 
These large dE/dX exceed
dE/dX of the IP muons crossing Si layers of VXD and Tracker, see 
 Figures~\ref{fig:dedx_vs_ekin_hits_e_from_mars_g_n}-\ref{fig:dedx_vs_ekin_hits_ip_mu_primary}. One can expect a modest improvement in rejection factor if 
using likelihood-ratio test instead of just threshold.

\subsection{Double Layer Criteria}
A stacked layer design to reduce muon collider random neutral background 
occupancy based on inter-layer correlation in the silicon detector was introduced
in \cite{Geer}. A single layer was replaced with two sublayers being 1-2mm 
apart and located in magnetic field (B$\sim$4T). The soft tracks from the muon 
collider background hits in one sublayer do not reach the second sublayer 
while IP physics charged tracks produce hits in both sublayers. Making readout 
of appropriate silicon pixels in both sublayers will suppress random background 
hits.

\begin{figure}[h]
\begin{minipage}{.45\textwidth}
  \centering
  \includegraphics[height=45mm]{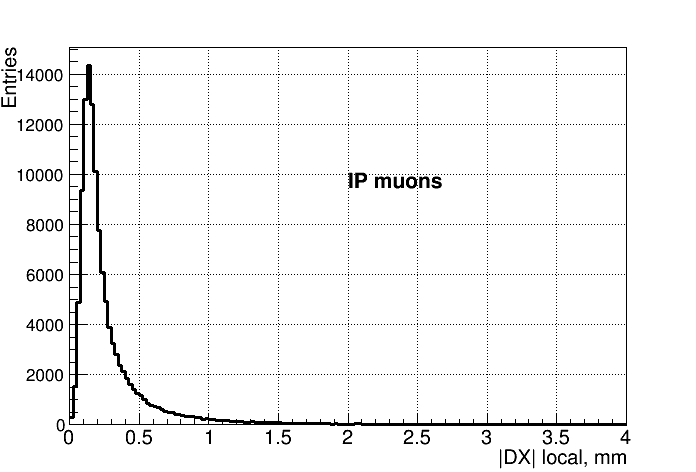}
  \caption{|DX| local for IP muon hits}
  \label{fig:IP_DXloc_abs_layer_18_1mm_1mfs}
\end{minipage}\hfill
\begin{minipage}{.45\textwidth}
  \centering
  \includegraphics[height=45mm]{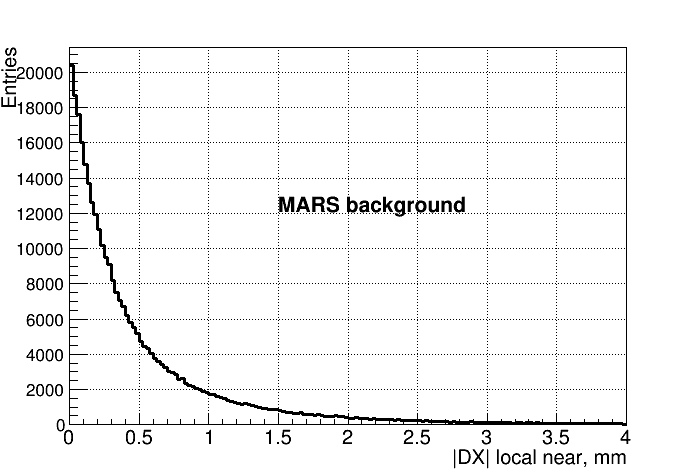}
  \caption{|DX| local between nearest cluster hits for MARS background}
  \label{fig:MARS_DXlocnear_abs_layer_18_1mm_1mfs}
\end{minipage}
\end{figure}

\begin{figure}[h]
\begin{minipage}{.45\textwidth}
  \centering
  \includegraphics[height=45mm]{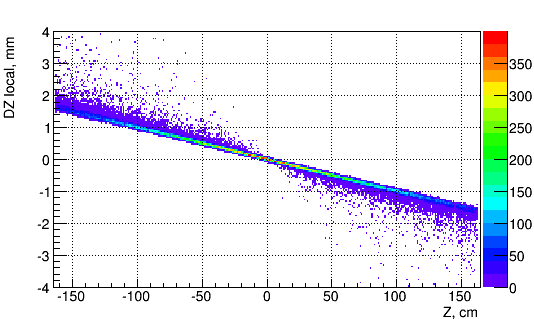}
  \caption{DZ local vs. Z global for IP muon hits}
  \label{fig:IP_DZloc_vs_zglob_layer_18_1mm_1mfs}
\end{minipage}\hfill
\begin{minipage}{.45\textwidth}
  \centering
  \includegraphics[height=45mm]{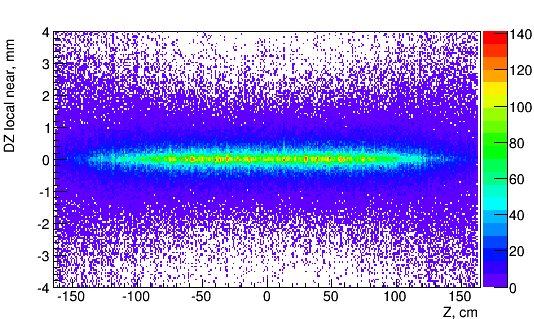}
  \caption{DZ local between nearest cluster hits vs. Z global for MARS background}
  \label{fig:MARS_DZlocnear_vs_zglob_layer_18_1mm_1mfs}
\end{minipage}
\end{figure}

In analysis we used $\sim$ 97\% IP muon cluster hits efficient cuts on difference of the hit clusters local X and Z
coordinates in both sublayers of the given layer.  The coordinates were 
smeared with 
$\sigma_{res}$ = 6$\mu$m for the VXD and $\sigma_{res}$ = 15$\mu$m for the Tracker.
Figures~\ref{fig:IP_DXloc_abs_layer_18_1mm_1mfs} and \ref{fig:MARS_DXlocnear_abs_layer_18_1mm_1mfs} present distributions of absolute value of 
difference in X ($\mid$DX$\mid$) for IP muons and MARS background cluster hits in the 
outermost Tracker barrel layer in geometry with a 1 mm space between sublayers 
and a 3.5\,T magnetic field. To illustrate the $\mid$DX$\mid$ difference 
distribution 
for MARS background cluster hits the nearest cluster hit X local coordinate 
was used. 
The distributions of DZ local coordinates difference vs. global Z coordinate
 of the cluster hit in the outermost tracker barrel layer in the same geometry 
 for IP muons (Figure~\ref{fig:IP_DZloc_vs_zglob_layer_18_1mm_1mfs})
and MARS background 
(Figure~\ref{fig:MARS_DZlocnear_vs_zglob_layer_18_1mm_1mfs}) suggest implementation of two-sided cuts depending on Z and VXD (Tracker) barrel layer. 

\section{Results for IP Muon Efficiency and MARS Background Surviving Fraction}
\label{sec:sec5}
\begin{figure}[h]
\begin{minipage}{.45\textwidth}
  \centering
  \includegraphics[height=48mm]{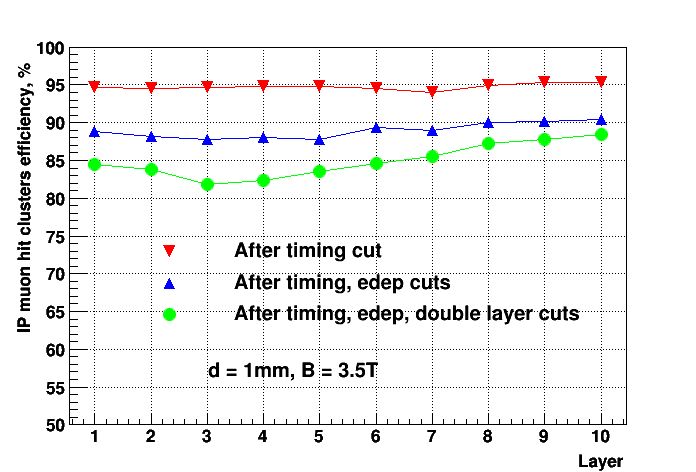}
  \caption{IP muon hit clusters efficiency vs. layers}
  \label{fig:ip_track_eff_1mm_1mfs}
\end{minipage}\hfill
\begin{minipage}{.45\textwidth}
  \centering
  \includegraphics[height=48mm]{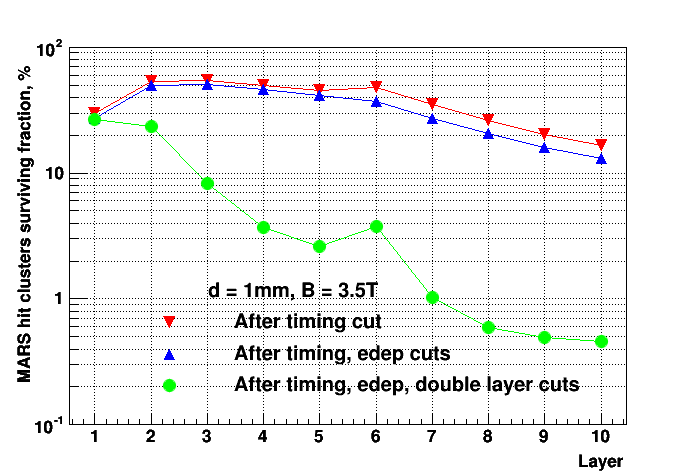}
  \caption{MARS hit clusters surviving fraction per sublayer vs. cuts and layers}
  \label{fig:surv_fract_1mm_1mfs}
\end{minipage}
\end{figure}

The IP muon cluster hit efficiency per layer after cuts is presented on 
Figure~\ref{fig:ip_track_eff_1mm_1mfs}. Here layers 1-5 are VXD barrel layers 
and 6-10 are Tracker barrel layers. The overall efficiency after all cuts is
$\sim$80-90\%. MARS background hit clusters surviving fraction per sublayer 
depends on the cuts and the layer. See Figure~\ref{fig:surv_fract_1mm_1mfs} for 
the geometry with 1 mm space between sublayers and a 3.5\,T magnetic field. Most 
of the rejection comes from timing and double layer cuts with an overall rejection 
factor as high as $\sim$200 in the outermost layers of the Tracker barrel. 
Such high suppression of background is due to low hit clusters density in 
these layers where the double layer criteria becomes the most powerful.

\begin{figure}[h]
\centering
\includegraphics[height=50mm]{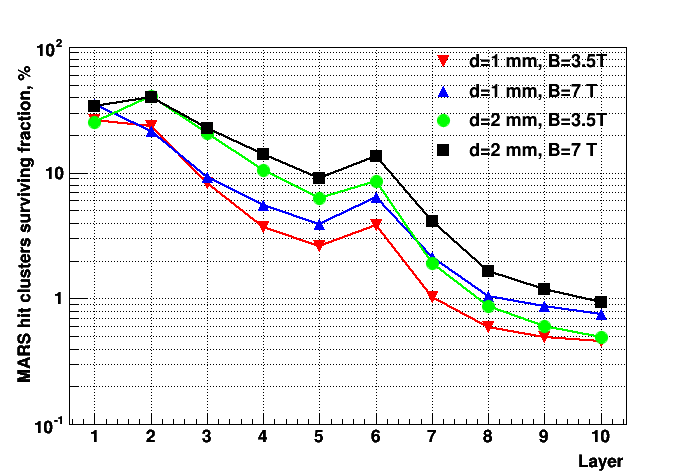}
\caption{MARS hit clusters surviving fraction per sublayer after all cuts in different geometries}
\label{fig:surv_MARS_vs_layer_all_geom}
\end{figure}

Figure~\ref{fig:surv_MARS_vs_layer_all_geom} presents the MARS hit clusters
surviving fraction after all cuts in different geometries. The background 
surviving fraction goes up with increasing sublayer space and magnetic field 
due to loosening double layer cuts (if the IP efficiency is kept the same). 
The overall MARS background hit clusters surviving fraction in barrel VXD and 
Tracker sublayers is $\sim$3\% in 1 mm geometries and $\sim$4-5\% in 2 mm 
geometries at IP efficiency of $\sim$85\% per layer.

\begin{figure}[h]
\begin{minipage}{.45\textwidth}
  \centering
  \includegraphics[height=50mm]{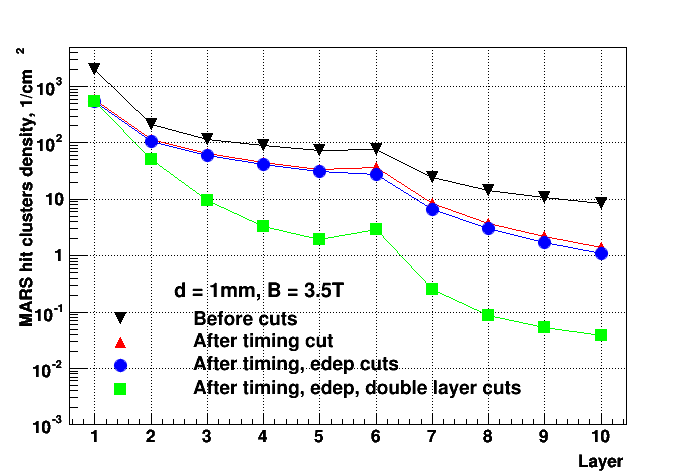}
  \caption{MARS hit clusters density per sublayer vs. cuts and layers}
  \label{fig:cluster_density_1mm_1mfs}
\end{minipage}\hfill
\begin{minipage}{.45\textwidth}
  \centering
  \includegraphics[height=50mm]{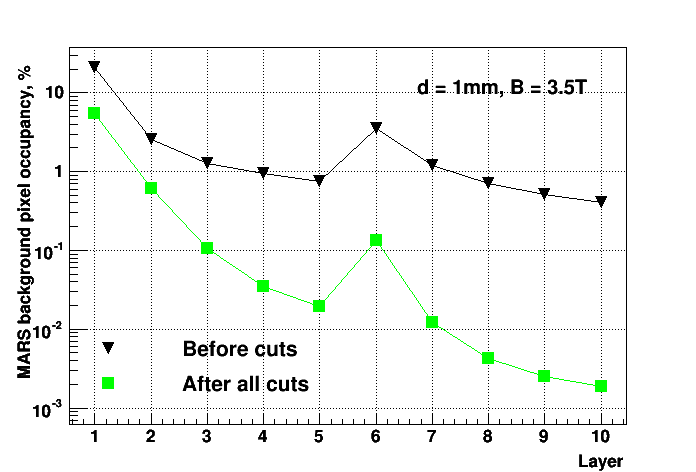}
  \caption{MARS background pixel occupancy per sublayer vs. layer}
  \label{fig:pixel_occup_1mm_1mfs}
\end{minipage}
\end{figure}

The density of MARS background hit clusters per sublayer in barrel layers of 
VXD and Tracker, before cuts and after cuts, is shown on  
Figure~\ref{fig:cluster_density_1mm_1mfs}. It remains high in the first two 
innermost barrel layers of VXD where the double layer cut is ineffective. 
The corresponding estimates of pixel occupancies are presented in 
Figure~\ref{fig:pixel_occup_1mm_1mfs} for 20$\times$20\,$\mu$m pixels in VXD barrel 
sublayers and  50$\times$50\,$\mu$m pixels in Tracker barrel layers. ILCroot provides 
module structure of the sensitive silicon sublayers. In analysis we used a simplified 
geometry of the pixels defined as sensitive sublayer of the module divided 
into square pixels. Only the pixels crossed by background track plus adjacent 
pixels were counted.

\section{Conclusion}
\label{sec:sec6}
The recent development in the design of the interaction region and 
machine-detector interface of the 1.5 TeV muon collider has demonstrated the
possibility of  suppression of muon beam background in the detector by 
more than three orders of magnitude. The ILCRoot simulation of the silicon vertex
and tracking detector hit response to the MARS15 background and the analysis of 
results on the hit level showed the feasibility of  
the use of a combination of timing, energy deposition and double layer criteria for 
further reduction of this background.
The timing criteria could be used in front-end electronics to decrease the
readout of background data. The next level of the background reduction
can be achieved by implementation of the energy deposition cuts and double layer
criteria in the analysis of the readout data (on a trigger level or in the tracking
algorithms).  

\acknowledgments
Work supported by Fermi Research Alliance, LLC under contract No. DE-AC02-07CH11359 with the U.S. Department of Energy through the DOE Muon Accelerator Program (MAP).

\end{document}